# Aspect ratio scaling of the Single Helical states in the Reversed Field Pinch plasmas

R. Paccagnella^

Consorzio RFX, Corso Stati Uniti 4, 35127 Padova, Italy

^and Istituto Gas Ionizzati del CNR

**Abstract:**

In this paper a new idea is presented which is able to predict the scaling law with the aspect ratio for the toroidal mode number of the so called Single Helicity states in Reversed Field Pinch plasmas. The model starts considering the pure electromagnetic interaction of an helical current with a conductive toroidal shell. The stability properties of the plasma become then essential in determining the emergence of the observed dominant mode.

**Introduction:**

The discovery of the so called "Single Helicity" (SH) states in the Reversed Field Pinch was made several years ago within the framework of 3D nonlinear cylindrical viscous-resistive magneto-hydro-dynamic (MHD) simulations of plasma dynamics [1,2]. Later, the existence of such states has been established also experimentally [3-8]. The main feature of the SH states in simulations [1,9] is the emergence of dominant quasi stationary mode with energy much higher than that of the concurrent helicities. The dominant helicity, in cylindrical numerical simulations, is generally not constant, since, for example, in some time windows a mode with toroidal mode number $n=-9$ can be dominant, in others the mode with other n's (-11,-12,-13) can take over [9]. The above mode numbers refer to a reference aspect ratio ( the ratio between the the major (R) and the minor (a) radius of the torus), R/a around 4. The dominant mode has instead always $m=1$ as poloidal mode number.

Hence, in cylindrical simulations, there is a certain variability of the dominant mode and this was generally linked to the nonlinear interactions in the system. In experiments instead a much lower variability is observed and it seems that the dominant mode is mainly related to the aspect ratio. Although this is not the only difference between cylindrical numerical simulations and experimental (and hence by definition toroidal) results, this paper will mainly address the question if a simple explanation could exist to explain the clear experimental scaling between the dominant mode in the

SH states and the torus aspect ratio. For example, for R/a around 4, the experimental dominant mode has n=-7 [8,10], and corresponds to the first on-axis resonant mode.

In experiments [11] and also in cylindrical numerical simulations [12,13] it has been shown that the application of helical boundary conditions for the normal magnetic field at the wall with the same helicity of the on-axis mode can excite it. In this way however any helicity could be excited in principle.

In experiments, however, the appearance of the SH states seems not connected to to the presence/application of an "enhanced" edge radial field with a specific helicity. The most clear case is maybe that of the Madison Symmetric Torus (MST) where a thick quasi ideal wall surrounds the plasma, minimizing field errors and the total normal field at the wall, and even in this case, clear SH states often with n=-5 (and more rarely with n=-6) have been observed [14,5]. Incidentally the torus aspect ratio R/a is 2.88 in this case.

The purpose of this paper will be therefore to show that a simple explanation could exist for the experimentally observed "selection rule" which robustly associates to each aspect ratio a different helicity of the dominant mode.

The paper is organized as follows: section 1 introduces the basic elements of the electromagnetic calculations that can explain why a certain helicity is preferred for a torus with a given aspect ratio, section 2 discusses some stability properties of the RFP plasmas, in section 3 a description of the observed SH states and the comparison with the theories is presented, finally discussion and conclusions are given.

1. **Electromagnetic coupling models**

A recent paper discussing a pure electromagnetic model [15] shows that a minimum of the energy of an helical surface current density distributed on a toroidal shell exists, as a function of the torus aspect ratio, for a certain current helical pitch.

Using a different approach, based on the virial theorem, older studies [16,17] have shown that an helical toroidal current with a certain helicity is subject to reduced electromagnetic stresses.

The main features of these two approaches are summarized below.

## 1.1 Magnetic energy of a toroidal distributed helical current

Following Ref. [15], assuming that a surface current $\mathbf{J} = (J_\theta, J_\phi)$ (where $\theta$ and $\phi$ are respectively the poloidal and toroidal angles) is distributed on a torus of major radius R and minor radius a, the magnetic energy of such a current can be calculated as:

$$W = W_\vartheta + W_\varphi = \frac{\mu_o}{8\pi} \int_{\partial V} \int_{\partial V} \frac{J_\varphi(r) J_\varphi(r') + J_\vartheta(r) J_\vartheta(r')}{|r-r'|} dS\, dS' \qquad (1)$$

where the integrals are over the torus surface. In Eq.(1) the cross term 2 ($J_\theta(r) J_\phi(r')$)) has been neglected because helices with different handedness should have the same energy on a homogeneous (no cuts or holes are present and the electrical conductivity is uniform) toroidal surface. For a current density distribution $\mathbf{J} = (J_{\theta o}/(1+(a/R)\cos(\theta)), J_{\phi o})$ where $J_{\theta o}$ and $J_{\phi o}$ are constants, W in Eq.(1) can be calculated (see [15] for details) for different torus aspect ratios. The current density components define an angle:

$$\tan(\alpha(\theta)) = \frac{J_\theta}{J_\varphi} = \frac{a\, n}{\rho\, m}$$

where n and m are the number of turns in the short and long way around the torus surface respectively, and $\rho = R + a \cos(\theta)$. It is possible to define the pitch angle $\alpha_o$ of the helix taking $\theta = +/- \pi/2$ i.e.

$$\tan(\alpha_o) = \frac{a\, n}{R\, m}$$

The main results of Ref.[15] (at least for the purpose of this paper) are that W has a clear minimum for n>0 (with m=1) as a function of (a/R) and that $\alpha_o$ slowly approaches $\pi/2$ for increasing n/m (see Fig.3 and 4 in [15]).

### 1.2 Minimal stresses on helical toroidal coils

Studying the minimal stress of an helical coil wired on a torus surface in Ref.[16,17]
it was shown that a simple relation exists between the coil pitch and the torus aspect ratio.
This relation is derived starting from the virial theorem expressed as:

$$W_B = \int \frac{B^2}{2\mu_o} dV = \int \sum_i \sigma_i\, dV \qquad (2)$$

where $W_B$ is the system magnetic energy and $\sigma_i$ represent (positive) tensile stresses which are necessary to store the magnetic energy within the given volume.

Dividing both sides of Eq.(2) by $W_B$ the stresses becomes normalized and the sum then equals unity.

In the coil circulates a current $\mathbf{I} = (I_\theta, I_\phi)$ that defines the pitch number of the coil as,

$N = I_\theta / I_\phi$.

With simple considerations the normalized stresses in poloidal and toroidal direction are expressed in terms of the pitch N of the helical current and of the torus aspect ratio $A = R/a$ as [16]:

$$\widehat{\sigma_\varphi} = \frac{A^2 \log 8A - A^2 - \frac{N^2}{2}}{A^2 \log 8A - 2A^2 + \frac{N^2}{2}}$$

$$\widehat{\sigma_\vartheta} = \frac{N^2 - A^2}{A^2 \log 8A - 2A^2 + \frac{N^2}{2}}$$

Assuming a uniform structure, the configuration with the minimum stresses is that in which poloidal and toroidal normalized stresses are equal to 0.5 [17].

This configuration is obtained when:

$$N^2 = \frac{2}{3} A^2 \log 8A \qquad (3)$$

2. **Reversed Field Pinch stability: the role of the on axis mode**

The Reversed Field Pinch (RFP) states, characterized by the spontaneous reversal of the toroidal field at the plasma edge, were obtained experimentally several years ago [18] and were interpreted in a simple and elegant way as relaxed states that correspond to a flat current density profile [19]:

$$\vec{J} = \nabla \times \vec{B} = \mu \vec{B} \qquad (4)$$

where $\mu$ is a constant (related to the current and the toroidal magnetic flux of the plasma).

Although Eq.(4) and the relaxation theory behind it, has been very successful in describing many important features of the experimental results, there are important aspects that are missed by Eq.[4] (for constant µ).

One of these aspects is the role of the on axis mode on the stability of the RFP devices.

In cylindrical geometry the solution of Eq.(4) is obtained in terms of Bessel Functions of the first kind, and the relaxed state results axi-symmetric when µ a < 3.11 (a being the cylinder minor radius) (and also ideally stable [20] ). For µ a = 3.11 the minimum energy state corresponds instead to a non-axisymmetric solution with m=1 and k a =1.234 (m, k being the poloidal and axial mode numbers) [19].

Above the limit µ a = 3.176 the first mode to become unstable is that with m=1, k a=0.86 (or m=-1, k a =-0.86) [20].

The condition for the mode resonance is given by:

$$\vec{k} \cdot \vec{B} = \frac{m}{r} B_\vartheta + \frac{n}{R} B_\varphi = 0 \qquad (5)$$

(m and n has the same meaning that in section 1) and simply express the fact that the helix of the equilibrium field lines match that of the perturbation. Eq.(5) can also be rewritten as:

$$q = \frac{r B_\varphi}{R B_\vartheta} = -\frac{m}{n}$$

where q is the safety factor.

Therefore it can be seen that, since the Taylor's states have q(0)>0 (or q(0)<0 (depending on the choice of an arbitrary free constant that multiplies the Bessel Functions), the mode found in [20] is not resonant on axis, i.e. it doesn't match the field lines pitch near r=0. In fact the mode with m=1 and k=0.86 correspond to an instability which has its resonance in the vacuum region outside the plasma. Note that the non-axisymmetric minimum energy state obtained for µ a = 3.11 corresponds also to the pitch of the field lines in the vacuum region beyond the plasma edge.

The importance of this fact was already emphasized in [21] where a force-free configuration like that of Eq.(4) was studied but with the important difference that a spatial dependence was assumed for µ , i.e. µ = µ (r) with µ (r) = µ(0) (1-(r/a)$^\alpha$) , with α defining the degree of current peaking. This assumption can be justified assuming a partial relaxation of the system, in presence of a non-uniform electrical conductivity, which tends to suppress the current at the plasma edge. In [21] it was therefore observed the important role on the RFP stability (both ideal and resistive) of the on-axis resonant modes corresponding to: $q(0) = (2 a)/(3 R)$ and having $(m/n) < 0$.

The important feature that change the stability properties of the RFP is therefore the presence of a finite parallel current gradient on axis. In [21] and later in [22] it was clearly shown that the experimental data tend to lie near to the marginal stability curve of the internal resonant modes.

The growth rates of the on axis (or near axis) resonant modes is decided by the current profile peaking (the exponent $\alpha$) as shown in Fig.1 ( for two cases with a near to zero toroidal field at the wall ).

The n=-7 becomes unstable only for the more peaked case with $\alpha =3$ . In recent papers the effect of local jumps on the $\mu$ profile [23] or of the pressure [24] have also been found, beside the global current peaking, to be destabilizing for the on-axis modes.

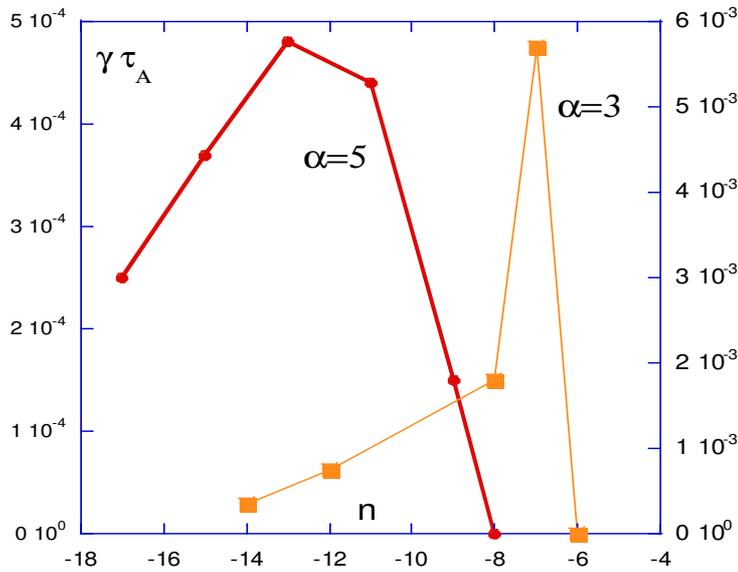

*Fig.1*: Growth rates (squares right and circles left y-axis) of resistive modes resonant on axis for two different current peaking values (with an ideal wall at r = 1.1 a and R/a=4.2 ).

### 3. Single Helical states: the aspect ratio effect

In Ref. [24] the phenomenology of the SH states in RFPs has been described in some detail and a model was constructed based on the stability of the on-axis mode to explain the observed experimental behavior, where the on-axis (or near axis) resonant mode becomes, time to time, the dominant one, it saturates to some amplitude and then it decays.

One thing that remains however unclear, as mentioned before, is why, contrary to cylindrical simulations, other modes with different helicities are not observed, and in all devices with different aspect ratios, most likely the on-axis resonant mode is instead seen detected.

The aspect ratios of existing RFP devices range ( fortunately ) over a large interval of values, from around 2 [25] to 6.8 [26].

We tried therefore to compare the prediction of the theories in section 1 for the pitch of the helical current as a function of the torus aspect ratio with the RFP experimental results (see Fig.2).

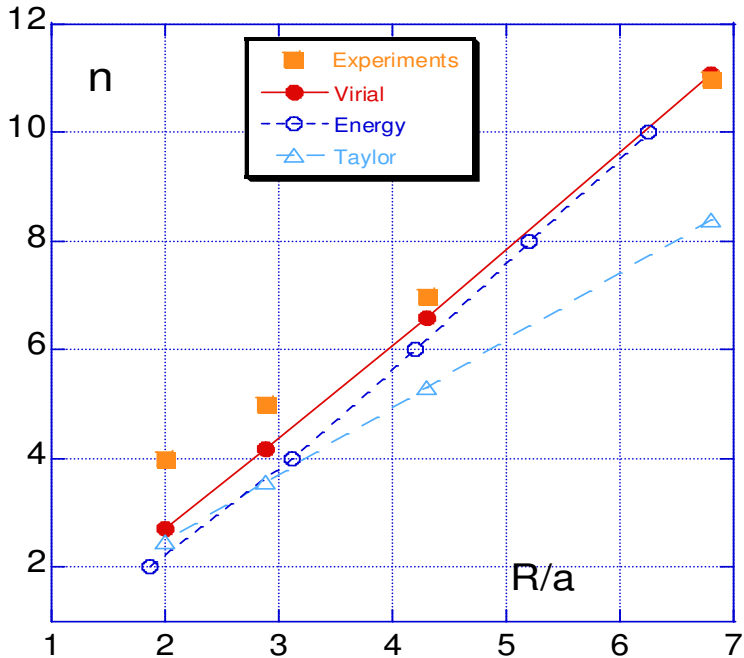

*Fig.2: Experimental points for the SH dominant mode (plain squares) compared with the virial theory (from Eq.(3) (plain circles) and minimum energy theory (as deduced from Ref. [15] Fig.4) (empty circles) and Taylor's prediction (triangles) vs. aspect ratio( for m=1 modes). (The n's values along the theoretical curves should be approximated to the next integer).*

As it can be seen the agreement with the simple electromagnetic theories described in Section 1 is very good. In particular the virial theorem theory [17] seems to catch through Eq.(3) the experimental trend very well. It should be noted that the experimental points deviate mostly from the theoretical prediction at low aspect ratio, where however the theory [17] is not precise since the poloidal current distribution on the torus is assumed uniform along the poloidal angle. This approximation becomes clearly less acceptable at low aspect ratio. It should also be noted that the points at the lowest aspect ratios ( R/a=2 and 2.88)  have some experimental  variability. For the aspect ratio 2 also the n=-3 was observed [25] with some probability, in better agreement with the virial theorem prediction, while for R/a=2.88 beside the n=-5 [14],  the n=-6 has been also detected [5] (in worse agreement with theory).

It is in any case remarkable that a simple equation like Eq.(3) is able to reproduce so well the experimental trend.

In Fig.2 also the prediction of the Taylor's relaxation theory is shown. In this case the curve corresponds to the line n= 1.234 (R/a) (see section 2) obtained in correspondence to the non axi-symmetric minimum energy solutions. As we already discussed these solutions correspond to the wrong helicity (resonant with the vacuum field outside the plasma). However it disagrees with the experimental data even for large R/a (where the theory instead is supposed to be perfectly applicable).

Regarding the variability of the dominant mode it should be noted that the results given in Fig.2 refer to the RFP operation at shallow reversal of the toroidal field. The parameter F  [19] , i.e. the ratio between the toroidal field at the wall and the average toroidal field over the plasma cross section, is for these cases typically  in the interval -0.1 < F < 0. As reported for example in [25] there is a dependence of the dominant mode number also on F (see discussion).

 **Discussion:**

As mentioned in section 1 according to the minimum energy or to the virial theories in presence of a homogeneous toroidal shell, preferred helicities with both handedness (positive and negative n's) are possible. In Fig.2  however the experimental points refer to negative n's since, as discussed in section 2 the on-axis resonant mode  is the one detected in experiments, due to  the parallel current profile within the plasma region and in particular to its gradient near the magnetic axis.

Once the plasma current profile has determined a group of unstable modes, we have shown that  the boundary can have an influence in determining the "selection" of a specific helicity, i.e. that which corresponds to the minimum stress and/or to the minimum energy of the wall current.

It seems therefore that the existence of a virtual or of a real good conducting structure favors the appearance of the spontaneous SH states. The condition that distinguishes an ideal shell from a penetrated (or resistive) one is that of having a zero normal component of the magnetic field. This condition, is physically, realized by toroidal and poloidal currents induced on the shell, which are able to counteract the magnetic field penetration. These are similar to the toroidal distributed shell localized currents that are considered by the theories briefly described in section 1.

In [10] for example, it was already noted that the virtual shell operation in RFX, i.e. a feedback scheme in which an array of active coils is used to mimic an ideal shell surrounding the plasma, clearly favored the emergence of the SH states. The same is true for all other RFP's with active feedback [25,26]. In devices like MST [5], where no active feedback system is present, a thick quasi-ideal shell surrounds the plasma and the SH states are also routinely observed.

We remind that also in Taylor's relaxation theory a fundamental role is played by the presence of an ideally conducting wall surrounding the plasma that acts as a flux conserver and also allows the global helicity invariance [19]. It could therefore be hypothesized that also for the appearance and sustainment of the SH states the presence of global invariants constitutes an essential ingredient.

The situation seems therefore, within our approach, different from experiments [11] or cylindrical numerical simulations [12,13], where a non-vanishing radial field with a given helicity is applied at the plasma boundary to excite the selected helicity. In our case it is exactly the opposite situation. In fact, the toroidal shell responds preferentially to specific modes and generates for them more efficiently the required back reacting radial field that is able to maintain an ideal shell boundary condition and to efficiently oppose the penetration of the radial field. This mechanism explains, at the same time, why other helicities are not observed, whereas, in the actively applied perturbations cases, in principle, every helicity could be forced.

It is, therefore, the existence of a quasi ideal casing surrounding the plasma which is able to explain the observed dominant helicity at different aspect ratios. Clearly in both cases, i.e. when a thick passive shell or when a feedback system is present, the condition that the radial field would be exactly zero at the shell is not perfectly realized in practice. For example, a feedback system tries to react to the sum of the plasma mode and of the shell currents in order to achieve a virtual shell condition. Due to the finite time response and circuitry limitations, generally, in this case, a residual radial field (of the order of 1-2 mT) is measured at the edge during the SH activity.

It is this small but finite mismatch with the perfectly conducting wall case that is able to create the electromagnetic linkage between the shell currents and the plasma modes, i.e the dominant mode varies in time trying to penetrate the shell, and the shell reacts to its electromagnetic forcing. The mechanism of selective mode excitation could be described therefore as follows: a group of modes are unstable in the plasma, they grow, the shell responds preferentially to a certain helicity, hence, currents with this specific helicity are generated on the shell, this in turn produces, due to the shell finite conductivity (presence of gaps or feedback systems limitations) a small, but not exactly zero, radial field with the same helicity that dynamically sustains the process, realizing most efficiently a quasi-ideal boundary for the specific helicity at each value of the aspect ratio.

Using ideas based on lumped parameters feedback models [27] it is easy to understand, at least qualitatively, that within our theory a good coupling between the mode and the wall leads to a more efficient stabilization of it. In fact starting from Eq.(20) of Ref[27] (reported here for reader's convenience only) as:

$$(1 - C)v^2 + \gamma_d \Delta (1 - C)v - \gamma_d \gamma_w = 0 \qquad (6)$$

where $v$ is the growth rate of the mode (under feedback), $C = (M_{pw} M_{wp})/(L_p L_w)$, $\gamma_d$ is the mode growth rate without wall (typically of the order of the inverse Alfven time), $\gamma_w = R_w/L_w$ is the wall inverse penetration time ($R_w$ and $L_w$ are the resistance and inductance of the wall) and

$$\Delta = \left(\frac{C + \frac{\gamma_w}{\gamma_d}}{1 - C}\right) - s, \quad s \text{ being the Laplace transformed variable.}$$

The important parameter, for our discussion is C, i.e. the coupling parameter between plasma and wall, where the M's are the mutual inductances. The parameter C varies in the range 0 to 1, being 1 when the plasma wall coupling is maximum. It can be easily verified that by increasing C the growth rate, obtained as the solution of Eq.(6) (by keeping the other parameters constant) decreases.

Physically this simply means that the wall is a more efficient stabilizer, i.e. the SH mode is efficiently stabilized by the wall (since their mutual inductances are high) (and also, as a consequence, by the feedback system). Therefore it should be concluded, as done in [23,24], that there are different reasons, possibly related to local effects (pressure or current gradients) near the mode resonance surface that act as destabilizing mechanisms at the origin of the observed mode growth and their cyclical behavior [24].

A slightly alternative, possibly more intuitive, model to estimate the radial field at the plasma edge is given by the following system:

$$\begin{aligned}\frac{\partial b}{\partial t} &= \gamma_d \, b \left(1 - c \, \frac{b}{b_{th}}\right) \\ \frac{\partial b_w}{\partial t} &= k \, \gamma_w \, b(t)\end{aligned} \qquad (7)$$

where $\gamma_d$ and $\gamma_w$ have the same meaning as before while k, c are constants. The first equation represents a growing mode [24] which saturates in time to a value $b_{th}$. b(t) can be considered the amplitude of a generic resistive mode near its resonant surface. The second equation links the edge value, $b_w$, to the value at the resonance through a coupling constant k ( with k< 1 representing qualitatively the radial decay of the mode) while its value is also influenced by the wall penetration time ($\gamma_w$=0 for an ideal wall). Using this simple model the results of figure 3 are obtained. It can be seen that for a given growth rate of the mode the ratio ($b_w$ /b) increases by increasing $\gamma_w$ i.e. by decreasing the wall penetration time.

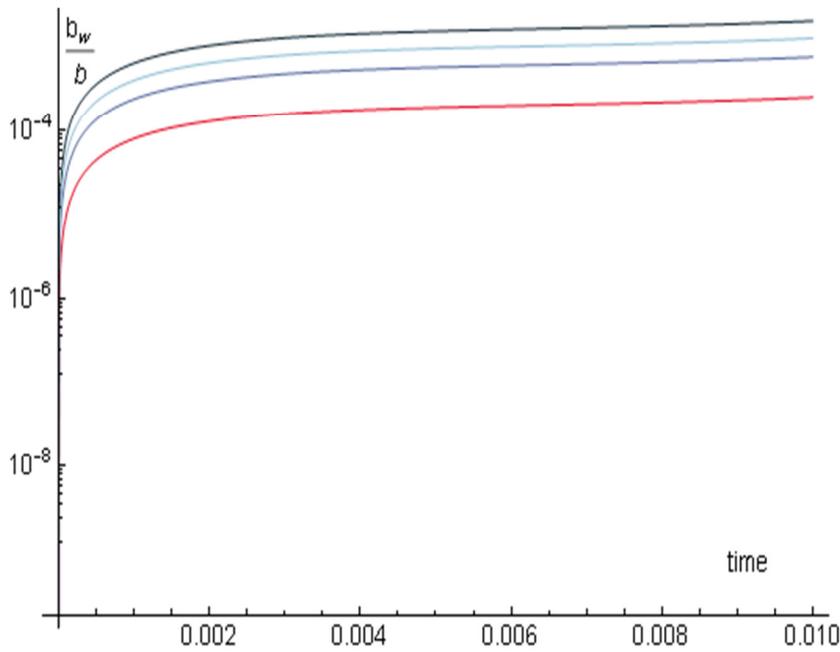

*Fig.3: $b_w$/b ratio vs. time (in ms) for $\gamma_d$ =500 $s^{-1}$ and $\gamma_w$ = 10,30,50,80 $s^{-1}$ (roughly from 100 to 10 ms wall penetration times) from the bottom to the top curve.*

Therefore we can conclude again that, being well shielded by the wall currents the SH mode will maintain a small value of ($b_w$/b) during its time evolution. This result is clearly different from what seen in cylindrical simulations with an externally applied perturbation [28], where a stronger edge perturbation is needed to excite the n=-7 on-axis mode with respect to other helicities. In our case, the good coupling between the on axis mode (n=-7 for R/a around 4) and the wall currents will maintain the radial field ( normalized to its value at the resonance surface) at the edge near to a minimum value.

From Eq.(7) it is also clear that a smaller growth rate of the mode will keep ($b_w$/b) lower and therefore will approximate better the ideal wall boundary condition. Clearly the growth rate of a resistive mode should depend inversely on temperature. Consistently, in experiments, it is observed that the SH states are favored at high temperature while cold high density plasmas tend to suppress them [29].

A final comment is necessary regarding the dependence, observed in experiments, of the dominant helicity at various values of the reversal parameter, F. For example in [25] for F > -0.5 n=-3-4 while for F<-0.5 n=-4,-5 are detected. In [30] a somehow stronger dependence is reported with variations from n=-7 at shallow F to n=-10 for F around -0.6.

On one hand it should be said that the purity and duration of the SH states have not been well discussed in literature for cases different from the shallow reversal case. This is a very important point, since it is generally observed that higher n's could emerge at deeper F but with shorter duration and purity, i.e. with a less clear energy gap of the dominant from the not dominant modes. This is also consistent with the observation that higher n's could penetrate the shell more efficiently through the existing cuts and holes, and therefore they are unable to efficiently excite reaction currents on the shell that can preserve the ideal wall boundary condition.

Second, it is clear, as already noted, that the emerging helicity is not only influenced by the shell response, but also by the plasma stability. It is a well established result (see for example [21]) that at deep reversal the maximum growth rate tends to move toward higher n's, while the growth rate of the on-axis mode generally decreases and eventually disappears. Lower F's (i.e. more deep reversed cases) correspond generally to a flatter current profile, and as seen in Fig,1, this shifts the plasma stability toward bigger n's, eventually stabilizing completely the on-axis mode. Generally, the turbulence level, and the related dynamo action and the system nonlinearities increase also for the deep reversed cases.

**Conclusion:**

The purpose of this paper was to answer to the simple question: why specific helicities are observed in SH states in RFP experiments with different aspect ratios?

A possible answer has been found, quite unexpectedly, in electromagnetic models which describe energetically favored helical currents on a toroidal conductive and homogeneous shell.

It was shown that an ideal shell can react preferentially to specific helicities. Therefore, it could be concluded that the dominant helicity in the SH states is the outcome between the plasma stability properties (i.e. the modes that are energetically favored in the plasma) and the energetically favored helicity of the toroidal shell currents. In fact, the shell has a "resonant" response to the helicities that are excited in the plasma in such a way that the shell "impedance" is minimum for some particular pitches depending on the aspect ratio.

For the on-axis resonant modes, unstable for relatively peaked current profiles (at shallow F) in RFP's, we have shown that the "resonant" response of a conducting toroidal shell surrounding the plasma to specific mode numbers, is very good over the whole range of aspect ratios of the existing RFP's, justifying the fact that in different devices the on-axis mode is easily excited and is identified, in practically all cases, as the dominant mode in the single helical states. The prompt, energetically favorite, reaction of the shell is therefore able to provide, efficiently, quasi-ideal wall boundary conditions to the SH mode, although the presence of a residual small radial field at the edge (with the same helicity) is unavoidable also in presence of a feedback system.

Within our theory, there is therefore no need to invoke complicate physical mechanisms

( like nonlinear mode coupling schemes etc.) to explain the experiments, in particular for the observed aspect ratio scaling.

We can hypothesize that the presence of a virtual casing surrounding the plasma could play a similar role as in Taylor's theory [19], by allowing the presence of global invariants in the system that favor the emergence of the SH states. This important issue is studied in a companion paper.

This work also suggests, as a possible new manner to enhance the purity of the SH mode in RFPs, by increasing, as much as possible, the mutual inductance matching between the wall and the dominant mode.

**Acknowledgments:** I thank my colleague Paolo Zanca for helpful discussions and suggestions.